\documentclass{ws-procs9x6}

\begin{document}

\def\vecsq{\stackrel{{}_{\rightarrow {\scriptstyle 2}}}}
\def\simlt{\stackrel{<}{{}_\sim}}
\def\simgt{\stackrel{>}{{}_\sim}}
\def\simm{\stackrel{\sim}{{}_{-}}}
\def\mkstop{m_{\tilde t, k}}
\def\mktop{m_{\tilde t, k}}
\newcommand\be{\begin{equation}}
\newcommand\ee{\end{equation}}
\newcommand\bea{\begin{eqnarray}}
\newcommand\eea{\end{eqnarray}}
\newcommand\ba{\begin{array}}
\newcommand\ea{\end{array}}

\newcommand{\gsim}{ \mathop{}_{\textstyle \sim}^{\textstyle >} }
\newcommand{\lsim}{ \mathop{}_{\textstyle \sim}^{\textstyle <} }
\newcommand{\vev}[1]{ \left\langle {#1} \right\rangle }
\newcommand{\bra}[1]{ \langle {#1} | }
\newcommand{\ket}[1]{ | {#1} \rangle }
\newcommand{\EV}{ {\rm eV} }
\newcommand{\KEV}{ {\rm keV} }
\newcommand{\MEV}{ {\rm MeV} }
\newcommand{\GEV}{ {\rm GeV} }
\newcommand{\TEV}{ {\rm TeV} }

\title{SUPERSYMMETRIC HIGGS BOSONS IN A 5D ORBIFOLD MODEL}

\author{V.~DI CLEMENTE AND S.~F.~KING}

\address{Department of Physics and Astronomy, \\
University of Southampton, Southampton, SO17 1BJ, U.K. \\
E-mail: vicente@hep.phys.soton.ac.uk, sfk@hep.phys.soton.ac.uk}

\author{D.~A.~J.~RAYNER}

\address{Dipartimento di Fisica `G. Galilei',
Universit\'{a} di Padova and INFN, \\
Sezione di Padova, Via Marzola 8, I-35131 Padua, Italy\\
E-mail: rayner@pd.infn.it}


\maketitle

\abstracts{We analyze the phenomenology of the Higgs sector in a 5D model
compactified on an $S_1/Z_2$ orbifold with a compactification scale
$M_C \sim {\mathcal O}(TeV)$ where supersymmetry breaking is localized
on a brane at one of the fixed points. We show that the conventional
MSSM Higgs boson mass bounds in 4D can be violated when we allow the
gauge sector, Higgs and third family multiplets to live in the fifth
extra dimension.} 

\section*{Introduction}

Extra-dimensional supersymmetric models with a TeV compactification scale 
$M_C$ offer an exciting new environment for investigating electroweak
symmetry breaking (EWSB)~\cite{ewsb}.
Higher-dimensional fields can be equivalently described as infinite towers of
Kaluza-Klein (KK) resonances and the $k^{th}$ KK mode has a mass
$m_{k} \sim {\mathcal O}(k/R)$, where $R = M_{C}^{-1}$ is the compactification
radius of the extra dimension(s)~\footnote{Note that ${\mathcal N}=1$
SUSY in 5D is equivalent to ${\mathcal N}=2$ SUSY in 4D, and so 5D
hypermultiplets can be decomposed into an ${\mathcal N}=1$ multiplet plus
its mirror-conjugate.}.  These models have the attractive feature that
loops of virtual top and stop KK modes yield a negative contribution
to the Higgs mass, which can trigger EWSB around
the TeV scale.  Remarkably, the one-loop effective potential is found
to be finite due to ${\mathcal N}=1$ SUSY in 5D which implies that the Higgs
physics is completely independent of the high-energy physics above
some cutoff scale. 

\section*{Our Model}  \label{sec:setup}

\begin{figure}[th]
\centerline{\epsfxsize=4.5in\epsfbox{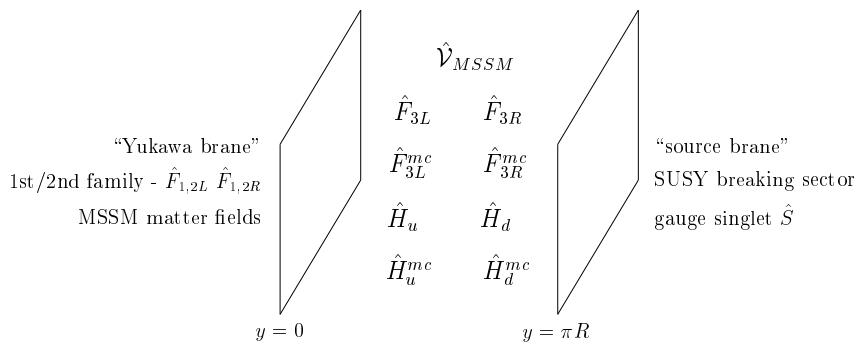}}   
  \caption{{\small The allocation of states in our 5D
  model is motivated by an explicit type I string construction \, $^{3}$.}}
  \label{fig:setup}
\end{figure}

We will summarize the features of our 5D model~\cite{us} as
shown in Figure ~\ref{fig:setup}. We
localize 4D 3-branes at the two fixed points $y=0,\pi R$ arising
from the compactification of the extra dimension on an
$S^{1}/Z_{2}$ orbifold.  SUSY is broken on the ``source'' 
brane at $y=\pi R$ when a localized gauge field singlet ($\hat{S}$)
acquires a non-zero F-term vacuum expectation value (VEV).

The first two MSSM ($\hat{F}_{1,2}$) families are confined to the ``Yukawa'' 
brane at $y=0$ where we are forced to localize Yukawa couplings, 
while the third family ($\hat{F}_{3}$), MSSM gauge sector 
($\hat{\mathcal V}_{MSSM}$) and Higgs superfields ($\hat{H}_{u,d}$) live in 
the extra dimensional bulk along with their mirror-conjugate
partners.  Bulk fields acquire
tree-level soft parameters due to their direct coupling with the SUSY
breaking brane through non-renormalizable operators defined at $y=\pi
R$, while first and second family squarks are suppressed by the
separation between branes~\cite{gaugino,rs} which alleviates the
flavour-changing neutral current problem.  The electroweak symmetry is
broken when the zero-modes of the Higgs fields acquire non-vanishing
VEVs ($<h_{u}> , <h_{d}> \neq 0$).   

The localization of the Yukawa couplings and soft SUSY breaking masses
at the fixed points leads to complicated mixing between different KK
modes when we decompose the 5D fields~\footnote{The mixing strength is
related to the SUSY breaking VEVs, and viable radiative EWSB requires
a non-universality between VEVs, i.e. $F_{S,\tilde{t}} \gg F_{S,H}$.}.  In
Ref.~\cite{us}, we diagonalized the top/stop mass matrices to find the
field-dependent KK mass spectra for the top:
\be
 m_{t,k}[h_{u}] = \left| k M_{C} + \frac{M_{C}}{\pi}
   \arctan\left(\frac{y_{t} \, h_{u} \, \pi}{\sqrt{2} M_{C}}\right) \right| 
 \quad\quad  (k= -\infty, \dots, \infty)
\label{eq:topmass}
\ee
and the field-independent stop KK modes:
\be
 \mkstop^2 \approx \left(k + \frac{1}{2} \right)^2 M_C^2   \qquad\quad 
    (k= -\infty, \dots, \infty)   
\label{eq:stopmass} 
\ee  
where $y_{t}$ is the 4D Yukawa coupling and we identify the physical
top mass as the $k=0$ mode of Eq.~\ref{eq:topmass}.  We used
dimensional regularization and zeta-function regularization  
techniques~\cite{kubyshin} to evaluate the top/stop contributions to the 
1-loop effective potential, and found that the top 
and stop contributions are separately finite (and free of 
ultraviolet divergences) due to supersymmetric cancellations.
  
\section*{Results}   \label{sec:results}

We applied the usual minimization conditions to the 1-loop effective
potential to extract the physical Higgs mass spectrum
in terms of a universal soft Higgs mass input parameter
(or equivalently $m_{A^{0}}$) and $\tan\beta$.  The results are shown
in Figure \ref{fig:higgs} for two different values of $\tan\beta$
where fine-tuning arguments~\cite{finetune} restrict
our analysis to compactification scales $M_C \simlt 4$ TeV.  
We also compare the MSSM results~\cite{pdg} with our model for 
$\tan\beta=1.5$ and find that our model is {\it not} excluded by the LEP 
signal~\cite{LEP}:
\begin{figure}[th]
\centerline{\epsfxsize=4.5in\epsfbox{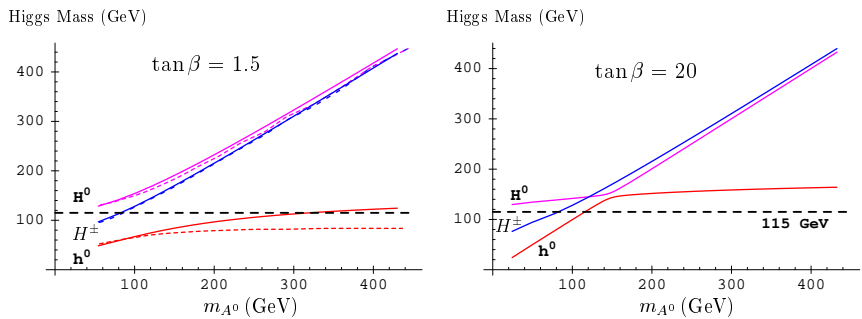}}   
\caption{\small The Higgs masses ($m_{h^{0}} , m_{H^{\pm}} ,
m_{H^{0}}$) against $m_{A^{0}}$ for $\tan\beta = 1.5$ (left  
panel) and $\tan\beta = 20$ (right panel).  The LEP candidate at 115 GeV
({\it bold dotted-line}) is shown for comparison with $m_{h^{0}}$.
For $\tan\beta=1.5$ (left panel), we also compare the MSSM results taken 
from ({\it dotted-lines}) against our model ({\it bold-lines}).}
  \label{fig:higgs}
\end{figure}


\begin{figure}[th]
\centerline{\epsfxsize=3.0in\epsfbox{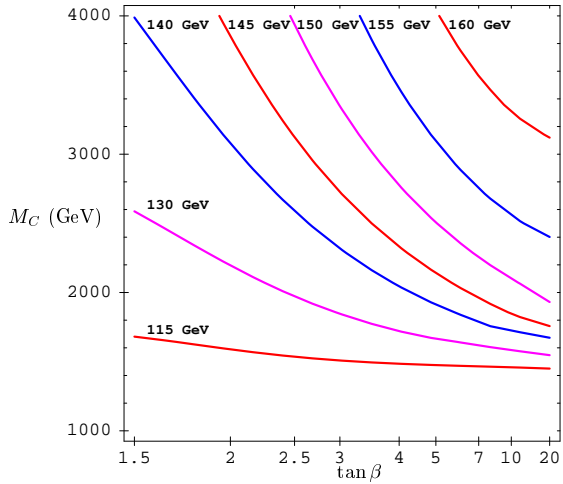}}   
\caption{{\small Mass contour plot for the lightest Higgs mass
$m_{h^0}$ as a function of the compactification scale $M_{C}$ (GeV) and
$\tan\beta \le 20$).}}
  \label{fig:mhcomp}
\end{figure}
 
In Figure \ref{fig:mhcomp}, we plot the lightest Higgs mass ($m_{h^{0}}$)
as a function of the compactification scale $M_{C}$ and $\tan\beta$.
The LEP data excludes the parameter space below the first contour at
$m_{h^{0}} = 115$ GeV ~\cite{LEP} which corresponds to a compactification 
scale $M_{C}  \approx 1.5 - 1.7$ TeV over the whole range of $\tan\beta$.
Notice that our model can easily accommodate the conventional 4D MSSM upper
bound on the lightest Higgs boson mass $m_{h^{0}} \sim 130$ GeV, and can be
pushed as high as $m_{h^{0}} \sim 160$ GeV with $M_{C} \approx 4$ TeV and
$5 \simlt \tan\beta \simlt 20$.  Remember that including additional matter
(e.g. gauge singlets in the NMSSM) can also raise the upper bound on
the lightest Higgs mass, but our ``minimal'' extension of the MSSM in
5D achieves higher mass bounds without adding extra matter content.


\end{document}